\documentclass[runningheads]{llncs}
\usepackage{graphicx}
\usepackage{enumitem}
\usepackage{tabularx}
\usepackage[hidelinks,bookmarks,bookmarksdepth=3,bookmarksopen]{hyperref}
\usepackage[colorinlistoftodos, textwidth=2.0cm]{todonotes}

\newcommand{\todofr}[1]{%
}

\title{Design and Implementation of a Remote Care Application Based on Microservice Architecture\thanks{This research is partially funded by the German Federal Ministry of Education and Research in the project ``QuartiersNETZ'' (grant number~02K12B061.)}}
\titlerunning{Design and Implementation of a Remote Care Application Based on MSA}

\author{Philip Nils Wizenty\and
	Florian Rademacher\and
	Jonas Sorgalla\and
	Sabine Sachweh
}

\authorrunning{Wizenty et al.}

\institute{
	University of Applied Sciences and Arts Dortmund,\\
	Institute for the Digital Transformation of Application and Living Domains,\\
	Otto-Hahn-Stra\ss{}e 23, 44227 Dortmund, Germany,\\	
	\email{\{philipnils.wizenty,florian.rademacher,jonas.sorgalla,sabine.sachweh\}\\@fh-dortmund.de}
}

\begin{document}
\maketitle

\begin{abstract}
Microservice Architecture (MSA) is an architectural style for service-based software systems. MSA puts a strong emphasis on high cohesion and loose coupling of the services that provide systems' functionalities. As a result of this, MSA-based software architectures exhibit increased scalability and extensibility, and facilitate the application of continuous integration techniques. This paper presents a case study of an MSA-based Remote Care Application (RCA) that allows caregivers to remotely access smart home devices. The goal of the RCA is to assist persons being cared in Activities of Daily Living. Employing MSA for the realization of the RCA yielded several lessons learned, e.g., (i) direct transferability of domain models based on Domain-driven Design; (ii) more efficient integration of features; (iii) speedup of feature delivery due to MSA facilitating automated deployment.

\end{abstract}

\section{Introduction}\label{sec:Introduction}
In the upcoming years, the number of people aged 60 and older will steadily increase and is predicted to worldwide grow from 901 million in 2015 to approximately more than 1.4 billions in 2030 \cite{Nations2015}. This will result in an increasing demand for caregivers which can not be covered by the labor market \cite{DonaldRedfoot2013}. Hence, new solutions are needed to cope with the resulting \textit{care gap}. The IT sector is one of the central domains from which such solutions are expected. That is, because it is perceived of being able to provide additional support for both persons being cared and caregivers by developing new supportive technologies \cite{DonaldRedfoot2013,Islam2015}. Next to mainly hardware-based solutions like wearables, Internet of Things based distributed software systems could aid in coping with the broadening care gap \cite{Islam2015}.

In this paper we present a distributed software system that addresses the domain of ambulant care. More specifically, the system denotes a Remote Care Application (RCA) \cite{RashidBashshur2009}, which leverages Microservice Architecture (MSA) \cite{Newman_2015} as its underlying architectural style. Its main purpose is to provide a platform for professional caregivers and nursing relatives to remotely interact with \textit{smart home devices} in households of persons being cared and hence support them in their Activities of Daily Living (ADL) \cite{6399501}, e.g., housekeeping, multimedia, taking medicine, or personal hygiene. We present the RCA in the form of a case study with the following objectives in mind: (i) provide MSA researchers with a practice-related, well-documented research object; (ii) present our experiences in MSA development; and (iii) elucidate our lessons learned in practical MSA development. Therefore, we describe the design, implementation and deployment of the RCA with a strong focus on architectural challenges and requirements.

The remainder of the paper is organized as follows. Section~\ref{sec:Requirements} identifies functional and non-functional requirements for the RCA. Section~\ref{sec:DesignAndImplementation} describes the RCA's design and implementation based on the requirement elicitation. Section~\ref{sec:Discussion} discusses the realized MSA solution for the RCA and the lessons learned during its implementation. Section~\ref{sec:Conclusion} concludes the paper.

\section{Functional and and Non-Functional Requirements of the Remote Care Application}\label{sec:Requirements}
This section describes the identified functional and non-functional requirements of the RCA. The requirements elicitation process was part of a holistic, iterative \textit{participatory design methodology}, which we developed and proved within the research project of which the RCA was one result \cite{Sorgalla2017PD}. Among others, the methodology comprises application-specific phases for stakeholder identification, as well as selection and application of the requirements elicitation process model being most appropriate for the stakeholders and the application to develop participatory. Employing the methodology resulted in (i) persons being cared, professional caregivers and nursing relatives being the relevant RCA stakeholders; (ii) the \textit{future workshop} method \cite{RobertJungk1996} in combination with \textit{goal-oriented requirements elicitation} \cite{Lamsweerde2001} being the most sensible process model for requirements elicitation. Tables~\ref{tab:FunctionalRequirements} and \ref{tab:Non-FunctionalRequirements} show the functional and non-functional requirement goals resulting from applying the process model. However, due to space constraints, we only list the tier1- and tier2-top-level goals and omit more fine-grained sub-goals.

\begin{table}
	\caption{Tier 1  and tier 2 functional requirement goals of the Remote Care Application}\label{tab:FunctionalRequirements}
	\begin{tabularx}{\textwidth}{c|>{\hsize=.2\hsize}X|>{\hsize=.7\hsize}X}
		ID&Goal&Description\\
		\hline
		T1-FG-1&Remote Support&The RCA must enable caregivers to remotely assist persons being cared in human-technology interaction scenarios.\\
		T1-FG-2&Preparing Household&The RCA must enable caregivers to remotely prepare a household in advance of an ambulant care visit.\\
		T2-FG-1&Smart Home&The RCA must be connected with various smart homes and be able to read device states.\\
		T2-FG-2&Remote Control&Devices in connected smart homes must be controllable.\\
		T2-FG-3&Access Rights&Device control must be explicitly granted by admins.\\								
	\end{tabularx}
\end{table}

The goals~T1-FG-1 and T1-FG-2 in Table~\ref{tab:FunctionalRequirements} describe the general purpose of the RCA. Accordingly, the RCA provides means to remotely assist household residents in human-technology interaction scenarios, e.g., programming the washing machine. Additionally, the RCA enables ambulant caregivers to dynamically prepare for their service remotely, e.g., heating up the bathroom in advance of their arrival. Based on tier 1, tier 2 Goal~T2-FG-1 addresses the need to read data from smart home devices with the RCA to, e.g., display current and historical device states. Next to basic read access, selected stakeholders may have to control devices, as expressed in Goals~T2-FG-2 and T2-FG-3.

\begin{table}
	\caption{Non-functional requirement goals of the Remote Care Application}\label{tab:Non-FunctionalRequirements}
	\begin{tabularx}{\textwidth}{c|>{\hsize=.15\hsize}X|>{\hsize=.75\hsize}X}
		ID&Goal&Description\\
		\hline
		NG-1&Scalability&The RCA must be able to handle thousands of households.\\
		NG-2&Security&Security must be high due to personal data being involved.\\
		NG-3&Availability&High availability on the basis of increased resilience, robustness and functional independence.\\
		NG-4&Extensibility&New functionalities need to be flexibly integrable and providable.\\
	\end{tabularx}
\end{table}

Table~\ref{tab:Non-FunctionalRequirements} shows non-functional requirement goals related to the RCA. They were actually the main drivers for implementing the RCA on the basis of MSA (cf. Subsection~\ref{subsec:MotivationForMSA}). Goal~NG-1 expresses the need for high scalability of the RCA being particularly relevant for professional caregivers having a potential customer base of thousands of households. Additionally, because of the high sensitivity of data related to the persons being cared and in case of misuse the potential of burglary, data and communication security must follow a high state of practice (goal~NG-2). The RCA must also exhibit a high degree of availability to enable quick reaction by professional caregivers and nursing relatives in cases of emergency (goal~NG-3). Another relevant concern is extensibility (goal~NG-4) to allow flexible integration of new smart home technologies and devices as well as new functionalities, e.g., means for data analysis, at runtime.

\section{Design and Implementation of the Remote Care Application}\label{sec:DesignAndImplementation}
This section describes the design and implementation of the RCA according to the identified requirements (cf. Section~\ref{sec:Requirements}). First, Subsection \ref{subsec:MotivationForMSA} elucidates our decision to apply MSA to the RCA w.r.t the requirements. Next, Subsection~\ref{subsec:ApplicationDesign} presents the application design. Finally, Subsection~\ref{subsec:Implementation} describes the RCA's implementation. 

\subsection{Motivation for Microservice Architecture}\label{subsec:MotivationForMSA}
We chose MSA as the RCA's underlying architectural style mainly because of the high requirements for scalability, availability and extensibility, i.e., goals~NG-1, NG-3 and NG-4 in Table~\ref{tab:Non-FunctionalRequirements}. The scalability requirement is satisfied because microservices are scalable independently due to functional isolation and technical self-containment \cite{Newman_2015}. Availability can be achieved because MSA fosters the definition of well-defined, functional service boundaries. Eventually, compared to monolithic applications, this results in an increased resilience as the faulty service (i) fails instead of the whole application; (ii) can be identified more effectively than in tightly coupled monoliths \cite{Newman_2015}. Extensibility is an inherent characteristic of service-based architectural styles \cite{Erl2005} and comes with well-partitioned service boundaries, which is one central characteristic of MSA \cite{Newman_2015}.

\subsection{Model-based Design of the Application}\label{subsec:ApplicationDesign}
The RCA's design process employed various types of models applicable to MSA for different design-specific concerns, i.e., capturing of domain concepts and microservice identification, interface modeling, and deployment modeling \cite{Rademacher2018}.

\begin{figure}
	\centering
	\includegraphics[scale=0.23,trim=1cm 1cm 1cm 1cm]{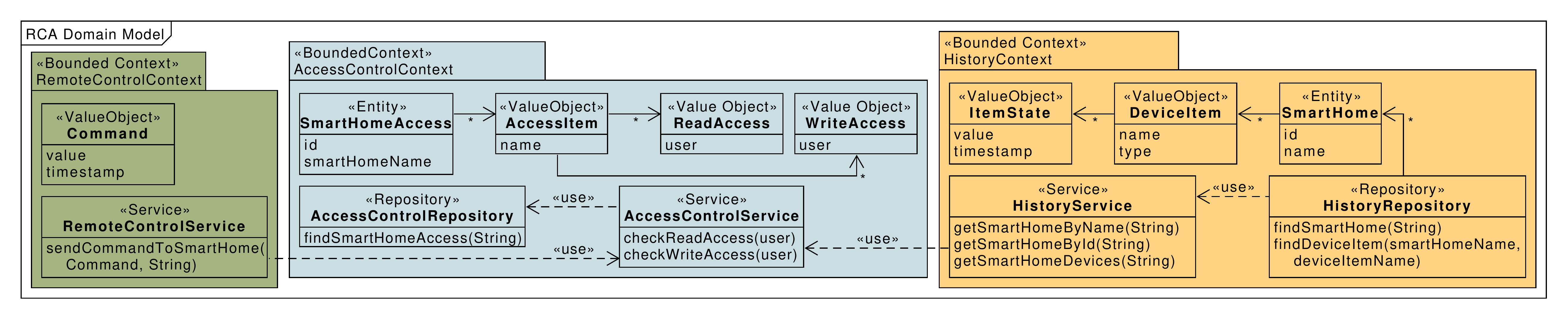}
	\caption{Domain model of the Remote Care Application}
	\label{fig:DomainModel}
\end{figure}

We applied the Domain-driven Design (DDD) methodology \cite{Evans2004} to iteratively capture relevant domain concepts and their relationships. This resulted in the \textit{domain model} shown in Figure~\ref{fig:DomainModel}, which we created in collaboration with domain experts, i.e., representatives of the RCA stakeholders (cf. Section~\ref{sec:Introduction}). Its notation and elements' semantics rely on a UML profile for DDD-based domain models \cite{Rademacher2017}, to prospectively enable semi-automatic model validation and code generation. Due to lack of space, however, we only present the result of the model creation process's final iteration. That is, a domain model with some technical information relevant to the RCA's implementation (cf. Subsection~\ref{subsec:Implementation}).

The domain model is decomposed into three Bounded Contexts, of which each denotes a candidate for a functional microservice \cite{Newman_2015,Rademacher2018}. The \texttt{HistoryContext} bundles domain-specific concepts that model data of connected smart homes, devices' components (called \textit{items}) and states. This corresponds to requirement goal~T2-FG-1 (cf. Table~\ref{tab:FunctionalRequirements}), whereby the current state of a given \texttt{DeviceItem} is the \texttt{DeviceState} with the highest \texttt{timestamp}. A concrete \texttt{SmartHome} instance may then, for instance, contain a \texttt{DeviceItem} ``ParlorLight\_Color'' with \texttt{DeviceState} \texttt{value} ``(210,0.25,1)'', which corresponds to light blue in the HSV color model. Furthermore, the context encapsulates a DDD Repository that models storage and retrieval of \texttt{SmartHome} instances and a Service to retrieve smart home and their device instances. The \texttt{RemoteControlContext} expresses characteristics for remote control of devices (cf.~T2-FG-2). The stakeholders regarded smart home devices as objects that may receive descriptive \texttt{Commands} like ``switch off palor light''. Therefore, the context's technical Service realizes communication of commands to a concrete smart home. The \texttt{AccessControlContext} models regulation of access rights (cf.~T2-FG-3). Within the collaborative domain modeling process we discussed several remote care control scenarios, that were identified in the future workshop with the stakeholders (see Section~\ref{sec:Requirements}). Together with the representative domain experts, we came to the conclusion that simple read and write access rules are sufficient. Central to the context is the \texttt{AccessControlService}, which is used by the other contexts and deals with checking the access rights of a given \texttt{user}. Hence, they are organized in \texttt{AccessItems} to subsume concrete, regulated domain concepts like \texttt{Command} or \texttt{DeviceItem}.

\begin{figure}
	\centering
	\includegraphics[scale=0.28,trim=1cm 1cm 1cm 1cm]{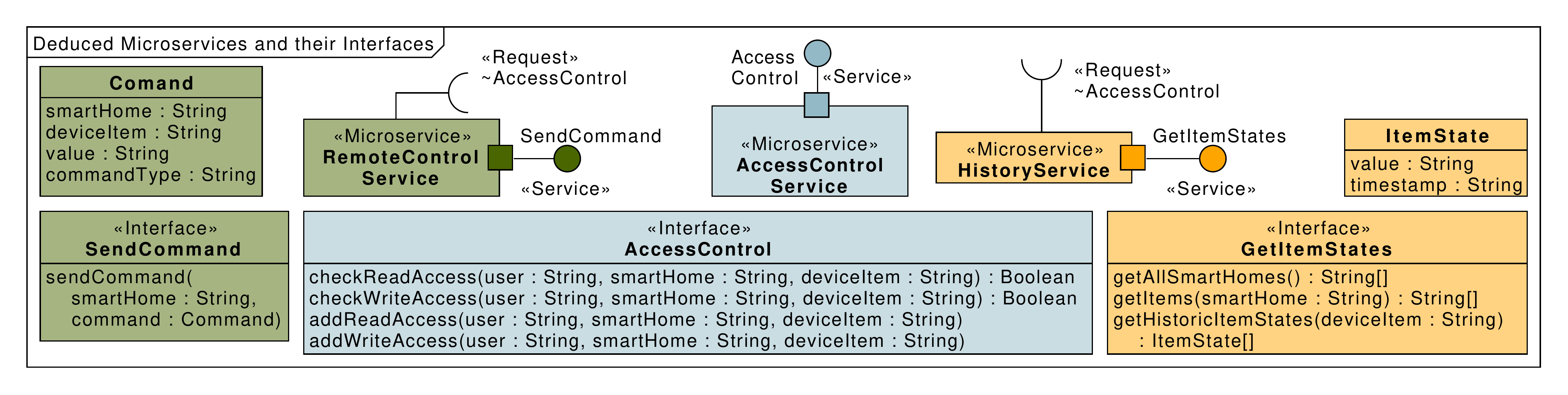}
	\caption{Interface model of the Remote Care Application in MSA-adapted notation \cite{Rademacher2018}}
	\label{fig:FunctionalServices}
\end{figure}

From the domain model, the \textit{interface model} \cite{Rademacher2018} depicted in Figure~\ref{fig:FunctionalServices} was deduced by mapping each Bounded Context in Figure~\ref{fig:DomainModel} to a microservice. As the interface model represents the technical implementation of the microservices, it needs to exhibit concrete technical information not present in the domain model, i.e., (i) interfaces with methods (deduced from the Bounded Contexts' Services of the domain model); (ii) types for exchanged data structures, e.g., \texttt{Command}, and their fields, e.g., \texttt{value}; (iii) service interaction relationships, e.g., the \texttt{HistoryService} consumes the \texttt{AccessControlService} (deduced from the \texttt{use} relationships in the domain model); (iv) additional elements necessary for the implementation, e.g., additional \texttt{add} methods in the \texttt{AccessControl} interface.

\subsection{Implementation}\label{subsec:Implementation}
Based on the interface model, we started to implement the RCA.  applied an MSA-specific, technically motivated Architectural Design (MSA-AD) \cite{Wizenty2017}. Next to the functional microservices as business-related components, it comprises common infrastructure components of MSA. 

Load Balancer and Circuit Breaker denote service-specific infrastructure components \cite{Wizenty2017}. Load Balancers may cope with increased amounts of requests by measuring incoming network traffic and distributing requests to different service instances. Circuit Breakers on the other hand increase the resilience of microservices by blocking requests that continuously result in errors or communication faults. Satisfying requirement goals~NG-1 and NG-3 (see Table~\ref{tab:Non-FunctionalRequirements}), both infrastructure components contribute to the RCA's scalability and availability. Another key infrastructure component of the MSA-AD relevant to the RCA is the Discovery Service \cite{Wizenty2017}. It allows functional and infrastructure microservices to expose their own interfaces, discover exposed interfaces of other services and establish a communication relationship with those services. The Discovery Services addresses requirement goal~NG-4 (cf. Table~\ref{tab:Non-FunctionalRequirements}) as it facilitates the flexible integration of new functionalities provided by microservices. The demand for security related to the RCA (goal~NG-2 in Table~\ref{tab:Non-FunctionalRequirements}) is satisfied by a dedicated Security Service \cite{Wizenty2017}. It acts as an identity provider for client authorization and authentication. The services employ token-based security by populating each request's data with an access token. The necessary user data is managed by the User Management Service. Hence, the caller can be clearly identified, even when the request is transitively delegated from service to service. In combination with an additional API Gateway Service, the Security Service realizes a Single Sign-On Gateway, i.e., a central point for authentication. Besides that, the API Gateway Service denotes the entry point to the RCA for external callers.

The implementation of the RCA is based on Spring Cloud\footnote{\url{http://projects.spring.io/spring-cloud}}. Consequently, each microservice is a standalone Java archive built with Spring Boot. The Discovery and API Gateway Service employ Eureka and Zuul, respectively. For architecture-internal communication synchronous RESTful HTTP is used. Furthermore, the API Gateway Service provides a REST endpoint for external requests. However, to increase the RCA's scalability (goal~NG-1 in Table~\ref{tab:Non-FunctionalRequirements}), asynchronous message-based communication via MQTT is applied with HiveMQ\footnote{\url{https://www.hivemq.com}} as message broker for receiving data from connected smart homes. Eventually, the Security Service was based on OAuth2\footnote{\url{https://oauth.net/2}}.

\begin{figure}
	\centering
	\includegraphics[scale=0.25,trim=1cm 1cm 1cm 1cm]{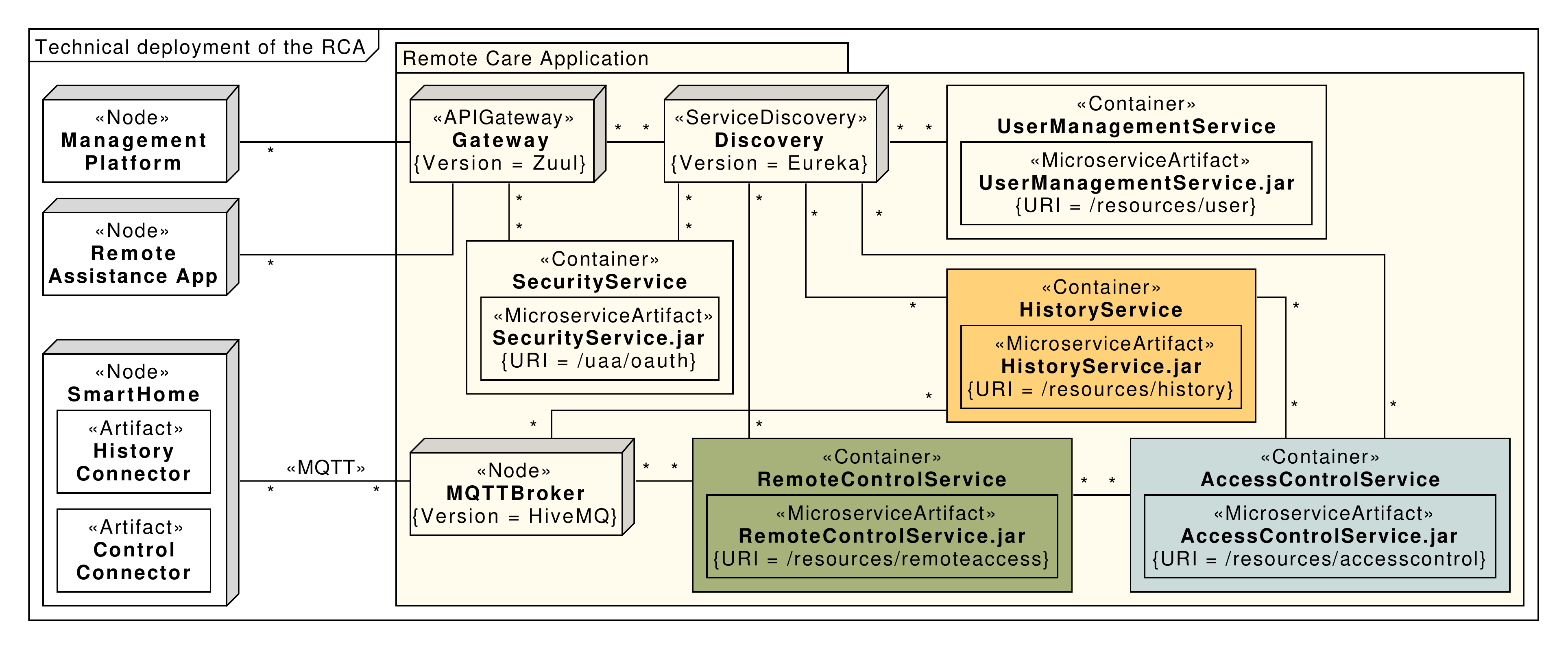}
	\caption{Deployment model of the RCA in MSA-adapted notation \cite{Rademacher2018}. It not otherwise stated, RESTful HTTP is used for communication purposes.}
	\label{fig:DeploymentModel}
\end{figure}

Figure~\ref{fig:DeploymentModel} presents the deployment overview of the RCA and connected software components. The RCA communicates with various external clients. Administrators can configure access rights and control a connected smart home with the \texttt{Man\-age\-ment\-Plat\-form} (cf. Subsection~\ref{subsec:ApplicationDesign}). It is mainly used on the office computers of professional caregivers. The \texttt{RemoteAssistanceApp} is the counterpart of the platform for mobile devices. Both control applications, i.e., platform and app, enable device control independent of a certain vendor. Additionally, the RCA acts as a mediator between smart homes and the control applications. Therefore, each \texttt{SmartHome} needs to execute an instance of Eclipse Smart Home\footnote{\url{https://www.eclipse.org/smarthome}}, because both \texttt{HistoryConnector} and \texttt{ControlConnector} are based on this framework to abstract from vendor-specific protocols. While the \texttt{HistoryConnector} transmits status changes to the RCA, the \texttt{ControlConnector} receives and executes remote control commands (cf. Subsection~\ref{subsec:ApplicationDesign}). 

To provide MSA researchers with an operating case study, whose creation was mainly driven by practical requirements, we made the RCA's code available as open source on GitHub\footnote{\url{https://github.com/SeelabFhdo/RemoteCareApplication}}.

\section{Discussion}\label{sec:Discussion}
This section discusses challenges we had to cope with in the RCA development.

First, we spent significantly more time with the engineering of infrastructure than functional microservices. The realization of the Security and API Gateway Service took the most time. The reasons for this are (i) a lack of documentation of the respective frameworks and (ii) the comparatively small business logic, i.e., while the RCA exhibits high degrees of scalability and extensibility its core functional capabilities are limited to a small number of microservices.

Second, the configuration of the development environment was partially cumbersome. By applying MSA, the application is decomposed into several small and autonomous services. This results in the physical development environment need to cope with a high amount of parallel processes, both functional and infrastructural. In our case this led to the regular development computers running out of resources when starting the RCA locally as a whole. Our solution was two-tiered. First we limited every microservices' available resources using JVM parameters. While thereby we were able to cope with the resource issue, this approach naturally reduced the application's performance and the opportunity to test scalability. Secondly, we distributed working services to a second development machine. Based on our experience we recommend to use a cloud server or multiple local machines directly in an early development stage.     

Third, we recognized that the functional microservices contained a lot of boilerplate code. Therefore, we developed a \textit{functional service template} \cite{Newman_2015}, i.e., a stub implementation that comprised all boilerplate code and needed to be filled only with the business logic. To be able to adapt this template to subsequent MSA projects in other domains, we developed a corresponding reusable build management tool that generates and embeds functional service stubs as well as stubs for infrastructure components \cite{Wizenty2017}. 

Fourth, the model-based design process (cf. Subsection~\ref{subsec:ApplicationDesign}) eased the implementation of the RCA. However, it also introduced an additional amount of effort as we deduced the interface and deployment model manually from the domain model \cite{Rademacher2018}. We expect that this extra work may be reduced by (semi-) automatically transforming the domain model into MSA-specific design models.

%\section{Related Work}\label{sec:RelatedWork}
%Domänenbeziehung
%Case Study im Microservice Bereich

\section{Conclusion}\label{sec:Conclusion}
In this paper, we presented a case study on how to apply MSA in the remote care domain. We identified requirements for the RCA leveraging participatory design techniques and employed DDD to create an appropriate domain model. From this, we deduced further design models which enabled us to derive the RCA's implementation consisting of several functional and infrastructure microservices. Hence, the case study may not only shed light on MSA's applicability for remote care, but also on how model-based design may aid in microservice development.

\bibliographystyle{splncs04}
\bibliography{literature}

\begin{thebibliography}{10}
\providecommand{\url}[1]{\texttt{#1}}
\providecommand{\urlprefix}{URL }
\providecommand{\doi}[1]{https://doi.org/#1}

\bibitem{Erl2005}
Erl, T.: Service-Oriented Architecture (SOA) Concepts, Technology and Design.
  Prentice Hall (2005)

\bibitem{Evans2004}
Evans, E.: Domain-Driven Design. Addison-Wesley (2004)

\bibitem{Islam2015}
Islam, S.M.R., Kwak, D., Kabir, M.H., Hossain, M., Kwak, K.S.: The internet of
  things for health care: A comprehensive survey. {IEEE} Access  \textbf{3},
  678--708 (2015)

\bibitem{Lamsweerde2001}
van Lamsweerde, A.: Goal-oriented requirements engineering: a guided tour. In:
  Proc. of the Fifth Int. Symp. on Requirements Engineering. pp. 249--262
  (2001)

\bibitem{Newman_2015}
Newmann, S.: Building Microservices. O'Reilly Media (2016)

\bibitem{Rademacher2017}
Rademacher, F., Sachweh, S., Z{\"u}ndorf, A.: Towards a uml profile for
  domain-driven design of microservice architectures. In: Software Engineering
  and Formal Methods. LNCS, vol. 10729, pp. 230--245. Springer (2018)

\bibitem{Rademacher2018}
Rademacher, F., Sorgalla, J., Sachweh, S.: Challenges of domain-driven
  microservice design: A model-driven perspective. IEEE Software  (May--June
  2018), in press

\bibitem{RashidBashshur2009}
Rashid~Bashshur, G.S.: History of telemedicine: Evolution, context, and
  transformation. New Rochelle (2009)

\bibitem{6399501}
Rashidi, P., Mihailidis, A.: A survey on ambient-assisted living tools for
  older adults. IEEE Journal of Biomedical and Health Informatics
  \textbf{17}(3),  579--590 (2013)

\bibitem{DonaldRedfoot2013}
Redfoot, D., Feinberg, L., Houser, A.: The aging of the baby boom and the
  growing care gap:a look at future declines in the availability of family
  caregivers. AARP Public Policy Institute  (2013)

\bibitem{RobertJungk1996}
Robert~Jungk, N.M.: Future Workshops: How to Create Desirable Futures.
  Institute for Social Inventions (1996)

\bibitem{Sorgalla2017PD}
Sorgalla, J., Schabsky, P., Sachweh, S., Grates, M., Heite, E.: Improving
  representativeness in participatory design processes with elderly. In: Proc.
  of the 2017 CHI Conf. Extended Abstracts on Human Factors in Computing
  Systems. pp. 2107--2114. ACM (2017)

\bibitem{Nations2015}
{United Nations}: World Population Ageing 2015. No. ST/ESA/SER.A/390 (2015)

\bibitem{Wizenty2017}
Wizenty, P., Sorgalla, J., Rademacher, F., Sachweh, S.: Magma: build
  management-based generation of microservice infrastructures. In: Proc. of the
  11th Europ. Conf. on Software Architecture (ECSA). pp. 61--65. ACM (2017)

\end{thebibliography}

\end{document}